# Photonic RF channelizer based on a 90 wavelength optical soliton crystal 49GHz Kerr microcomb

Xingyuan Xu, Mengxi Tan, Jiayang Wu, Andreas Boes, Thach G. Nguyen, Sai T. Chu, Brent E. Little, Roberto Morandotti, *Fellow, OSA, Senior Member, IEEE,* Arnan Mitchell, *Member, IEEE,* and David J. Moss, *Fellow, OSA, Fellow IEEE*

*Abstract*— **We report a broadband radio frequency (RF) channelizer with up to 92 channels using a coherent microcomb source. A soliton crystal microcomb, generated by a 49 GHz micro-ring resonator (MRR), is used as a multi-wavelength source. Due to its ultra-low comb spacing, up to 92 wavelengths are available in the C band, yielding a broad operation bandwidth. Another high-Q MRR is employed as a passive optical periodic filter to slice the RF spectrum with a high resolution of 121.4 MHz. We experimentally achieve an instantaneous RF operation bandwidth of 8.08 GHz and verify RF channelization up to 17.55 GHz via thermal tuning. Our approach is a significant step towards the monolithically integrated photonic RF receivers with reduced complexity, size, and unprecedented performance, which is important for wide RF applications ranging from broadband analog signal processing to digital-compatible signal detection.**

## I. INTRODUCTION

Receivers are key building blocks of modern RF systems including satellite communications, electronic warfare, and radar systems [1-10], as they can detect and analyze received RF signals. However, the instantaneous bandwidth of the received RF signal is generally beyond the capability of most analog-to-digital converters (ADCs). As such, to detect and analyze signals with powerful and flexible digital-domain tools, the broadband signal needs to be spectrally sliced into digital-compatible segments for separate digital processing [11] and this is achieved by RF channelizers. While electronic RF channelizer are subject to the bandwidth bottleneck, photonic approaches are promising since they can offer ultra-large bandwidths, low transmission loss and strong immunity to electromagnetic interference.

Significant effort has been directed towards photonic RF channelizers, which can be divided into two categories. The first relies on physically splitting and slicing the RF spectrum using a large number of spectrally dense and precisely centered narrow-linewidth filters [12-15]. While effective, this approach poses limitations in the overall footprint, resolution and operational bandwidth. The other category of photonic RF channelizers operates in a much more compact and efficient way by employing the Vernier effect between a multi-wavelength source and periodic optical filter to achieve a precisely stepped and broadband RF spectral channelization. Diverse types of devices and platforms have been used for this, such as stimulated Brillouin scattering [16-18], parametric processes in nonlinear fiber [19, 20], spectrally sliced incoherent sources [21], discrete laser arrays [22], electro-optic modulator-based frequency combs [11, 23, 24], and others [25, 26]. Yet those approaches face limitations in channel number, spectral resolution, and ease of being monolithically integrated, due to bulky discrete components.

Recently, microcombs [27-31], particularly those based on CMOS-compatible platforms [32-40], came into focus since they offer a large number of coherent wavelength channels in a $mm^2$-size footprint, and have enabled a wide range of RF applications [41, 42], such as RF true time delays [43, 44], transversal signal processors [45-50], frequency conversion [51], phase-encoded signal generators [52], and RF channelizers [53, 54]. Previously [54], we reported an RF channelizer based on a microcomb that achieved high performance in a compact footprint. In that approach, a 200 GHz spaced microcomb generated 20 wavelengths in the C band, which was combined with a passive MRR filter having an FSR of 48.9GHz. In that work, the RF signal was imprinted on the 200GHz comb which was then filtered with every $4^{th}$ resonance of the passive MRR. This mismatch between the free spectral ranges (FSR) of the microcomb and four times the passive MRR filter resulted in a frequency offset of > 4.4 GHz, equal to the frequency step between channelized RF spectral segments. This offset was much larger than the channel

X. Xu, M. Tan, J. Y. Wu, and D. J. Moss are with Optical Sciences Centre, Swinburne University of Technology, Hawthorn, VIC 3122, Australia. (Corresponding e-mail: dmoss@swin.edu.au).

A. Boes, T. G. Thach and A. Mitchell are with RMIT University, Melbourne, VIC 3001, Australia.

S. T. Chu is with Department of Physics and Material Science, City University of Hong Kong, Tat Chee Avenue, Hong Kong, China.

B. E. Little is with State Key Laboratory of Transient Optics and Photonics, Xi'an Institute of Optics and Precision Mechanics, Chinese Academy of Science, Xi'an, China.

R. Morandotti is with INSR-Énergie, Matériaux et Télécommunications, 1650 Boulevard Lionel-Boulet, Varennes, Québec, J3X 1S2, Canada, and is adjunct with the Institute of Fundamental and Frontier Sciences, University of Electronic Science and Technology of China, Chengdu 610054, China.

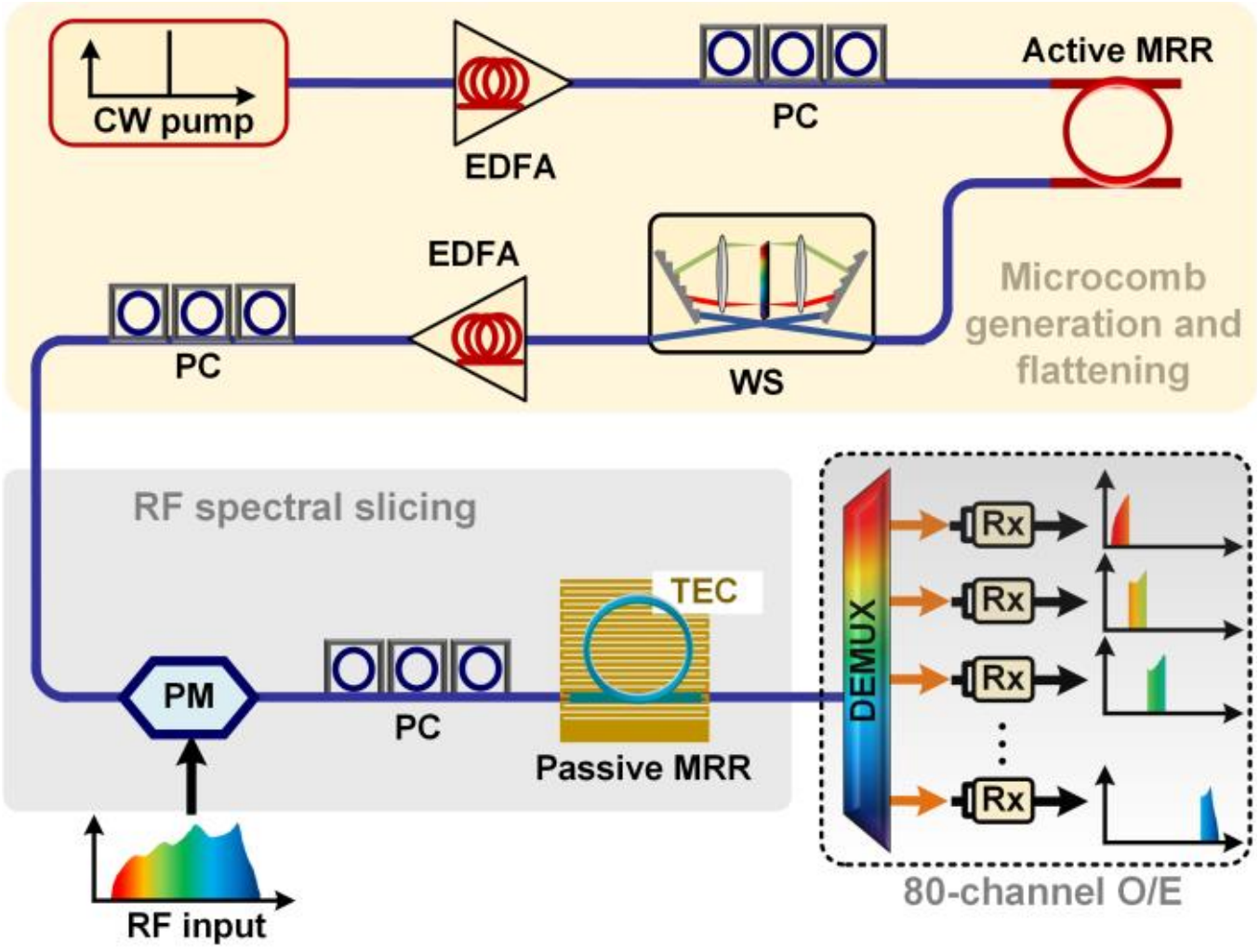

Fig. 1. Schematic diagram of the broadband RF channelizer based on a soliton crystal microcomb. EDFA: erbium-doped fibre amplifier. PC: polarization controller. MRR: micro-ring resonator. WS: Waveshaper. PM: phase modulator. TEC: temperature controller. DEMUX: de-multiplexer. Rx: Receiver.

resolution and so it was not possible to achieve continuous RF operational sampling bands, which resulted in a discontinuous RF operation band with only 4 channels operating simultaneously, limiting the overall instantaneous bandwidth – the product of the channel number and slicing resolution. The "gaps" were covered by thermally shifting the passive MRR. Finally the large FSR of the microcomb resulted in only 20 wavelength channels over the C-band.

Here, we report a photonic RF channelizer with greatly improved performance, achieved by using two MRRs with quasi-matched FSRs, both near 49GHz. The first MRR generates a soliton crystal microcomb while the second acts as a passive filter. This results in two significant benefits. First, the much smaller FSR of the comb source enables the generation of up to 92 wavelength channels in the C band. This leads to a significantly enhanced instantaneous operational bandwidth of 8.08 GHz [54]. By using an integrated high-Q MRR with a (roughly) matching FSR to slice the RF spectrum via the Vernier effect, a RF channelization step of 87.5 MHz is achieved. This leads to successful continuous channelization of the RF spectrum since the channelization step is smaller than the spectral resolution of 121.4 MHz. We also employ parallel phase-modulation to intensity-modulation (PM-IM) conversion across all wavelength channels — which directly offers an RF output in a compact and stable scheme without the need for separate local oscillator paths. Finally, the channelizer's operation frequency range can be dynamically tuned by adjusting the offset frequency between the microcomb and passive MRR. We thermally tune the passive MRR and verify successful RF channelization over a total RF range of 17.55 GHz. In addition to advanced RF performance, this approach offers a reduced footprint, lower complexity, and potentially lower cost.

## II. Operation Principle

Figure 1 shows the setup of the broadband RF channelizer that consists of three modules. The first module consists of the microcomb generation and flattening, where an active MRR is pumped by a continuous-wave (CW) laser to initiate parametric oscillation. With the MRR's high Q factor of over 1 million, the high nonlinear figure of merit, and tailored anomalous waveguide dispersion, sufficient parametric gain to generate Kerr frequency combs was provided. The state of the generated frequency comb is determined mainly by the detuning between the pump and the resonance, and the pump power. As such, by sweeping the pump wavelength from blue to red, diverse nonlinear dynamic states, including the coherent soliton states, could be triggered.

Assuming $N$ microcomb lines are generated with a spacing of $\delta_{OFC}$, the optical frequency of the $k_{th}$ ($k$=1, 2, 3, …, 92) comb line is denoted as

$$f_{OFC}(k)=f_{OFC}(1)+(k-1)\delta_{OFC} \tag{1}$$

where $f_{OFC}(1)$ is the frequency of the first comb line used on the red side. An optical spectral shaper (the commercially available Waveshaper) is then used to flatten the power of the comb lines to achieve equalized channel power.

In the second module, the flattened comb lines are directed to an electrooptical phase modulator, where the input broadband RF signal is multicast onto all the wavelength channels. Next, the replicated RF spectra are sliced by the passive MRR with an FSR of $\delta_{MRR}$, where the slicing resolution is given by the 3dB bandwidth of the resonator. As a result, the RF spectral segments on all wavelength channels are effectively channelized with a progressive RF centre frequency, given by

$$\begin{aligned} f_{RF}(k)&=f_{MRR}(k)-f_{OFC}(k) \\ &= [f_{MRR}(1)-f_{OFC}(1)]+(k-1)(\delta_{MRR}-\delta_{OFC}) \end{aligned} \tag{2}$$

where $f_{RF}$(k) is the $k_{th}$ channelized RF frequency, $f_{MRR}$(k) is the $k_{th}$ centre frequency of the filtering MRR. [$f_{MRR}$(1)– $f_{OFC}$(1)] denotes the relative spacing between the first comb line and adjacent filtering resonance, corresponding to the offset of the channelized RF frequencies, and ($\delta_{\mathrm{OFC}} - \delta_{\mathrm{MRR}}$) corresponds to the channelized RF frequency step between adjacent wavelength channels.

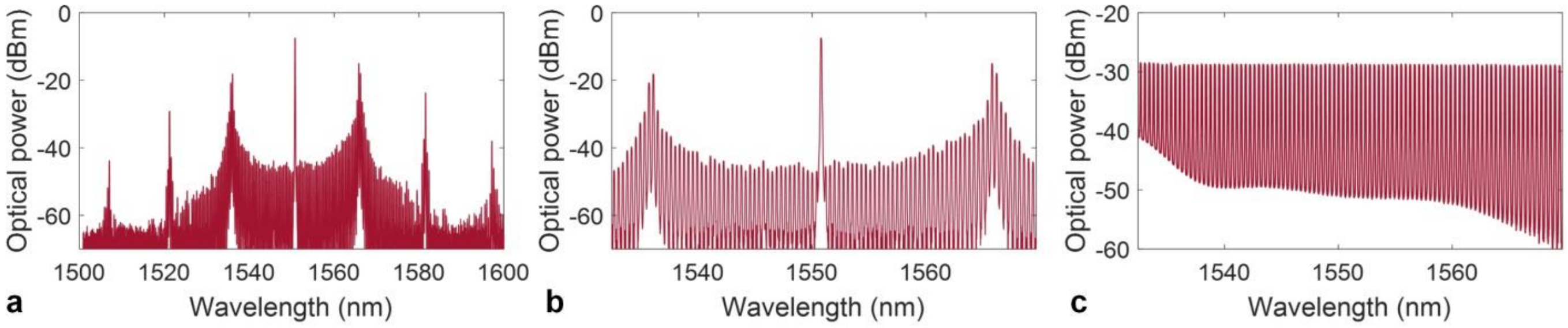

Fig. 2. Optical spectrum of the generated soliton crystal microcomb with (a) 100 nm and (b) 40 nm span. (c) Flattened 92 comb lines.

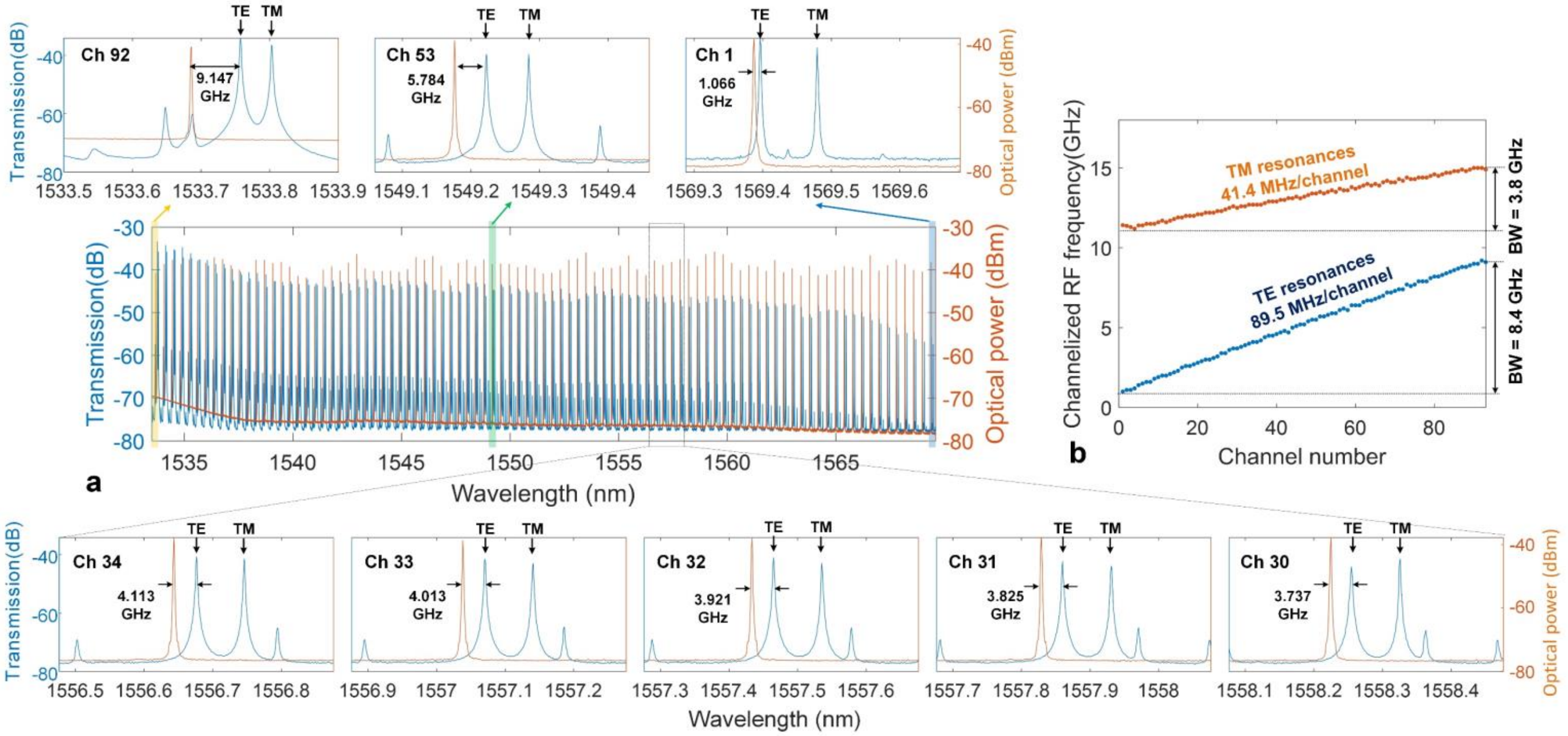

Fig. 3. (a) The measured optical spectrum of the micro-comb and drop-port transmission of passive MRR. (b) Extracted channelized RF frequencies of the 92 channels, calculated from the spacing between the comb lines and the passive resonances. Note that the labelled channelized RF frequencies in (a) is adopted from the accurate RF domain measurement using the Vector Network Analyzer, as shown in the next figure.

We employed phase modulation and notch filtering (i.e., the transmission of the passive MRR's through port) to achieve phase to intensity-modulation conversion. This approach does not require any other physical local oscillator paths to achieve coherent homodyne detection, and so is much more compact and stable than those involving interfering paths [54]. Finally, the wavelength channels are de-multiplexed and converted back into the electrical domain separately for ADCs and further digital domain processing.

## III. Experimental results

The active and passive MRRs were both fabricated in a CMOS-compatible doped silica glass platform, employing fabrication processes. First, high-index (the refractive index is ~1.66 at 1550 nm) doped silica glass films were deposited using standard plasma enhanced chemical vapour deposition, then photo-lithographically patterned and reactive ion etched to form waveguides with low surface roughness. Finally, silica glass (the refractive index is ~1.44 at 1550 nm) to form an upper cladding was deposited. This platform offers low linear loss (~0.06 dB·cm$^{-1}$), a moderate nonlinearity parameter (~233 W$^{-1}$·km$^{-1}$) and, in particular, a negligible nonlinear loss up to extremely high intensities (~25 GW·cm$^{-2}$) [39]. Thus, this has been an extremely successful platform for microcombs. The MRRs featured similar radii of ~ 592 µm and cross sections of 1.5×2 µm with Q factors > 1million and a through-port insertion loss of ~1.5 dB.

The active MRR was tailored to feature anomalous dispersion in the C band to enable parametric oscillation, together with a mode-crossing at ~1556 nm, which could initiate a background wave for soliton crystal generation. During comb generation, the pump power was boosted to ~ 2W with the wavelength swept manually from blue to red. As the detuning between the pump wavelength and the active MRR's resonance became small enough to ensure sufficient modulation-instability gain in the active MRR, primary combs could be initiated, followed by soliton crystal microcombs as the detuning was changed further. Distinctive 'palm-like' optical spectra of the soliton crystals were observed, as shown in Fig. 2 (a, b). The curtain-like spectrum was a result of the interference between the tightly-packaged solitons circulating along the ring cavity [55, 56]. The soliton crystal combs featured a spacing equal to the FSR of the active MRR, which was ~ 48.9 GHz, enabling up to 92 wavelength channels in the operation band of our equipment (the C-band Waveshaper). This category of microcomb is coherent and low-noise and, due to the ultra-high intracavity power, could easily be generated by manually sweeping the pump wavelength—much simpler that the generation of single-soliton states.

Next, the generated soliton crystal microcombs were flattened using an optical spectral shaper (Waveshaper 4000S) to equalize the power of the wavelength channels. We adopted a feedback control path to shape the comb lines' power accurately, which were monitored by an optical spectral analyzer and compared with the desired channel weights to generate an error signal used to program the loss characteristics of the Waveshaper. The flattened comb spectrum is shown in Fig. 2(c).

Then the 92 flattened microcomb lines were fed into a phase modulator and served as optical carriers, thus broadcasting the input RF signal onto all wavelength channels. The RF replicas were then spectrally sliced by a passive MRR.

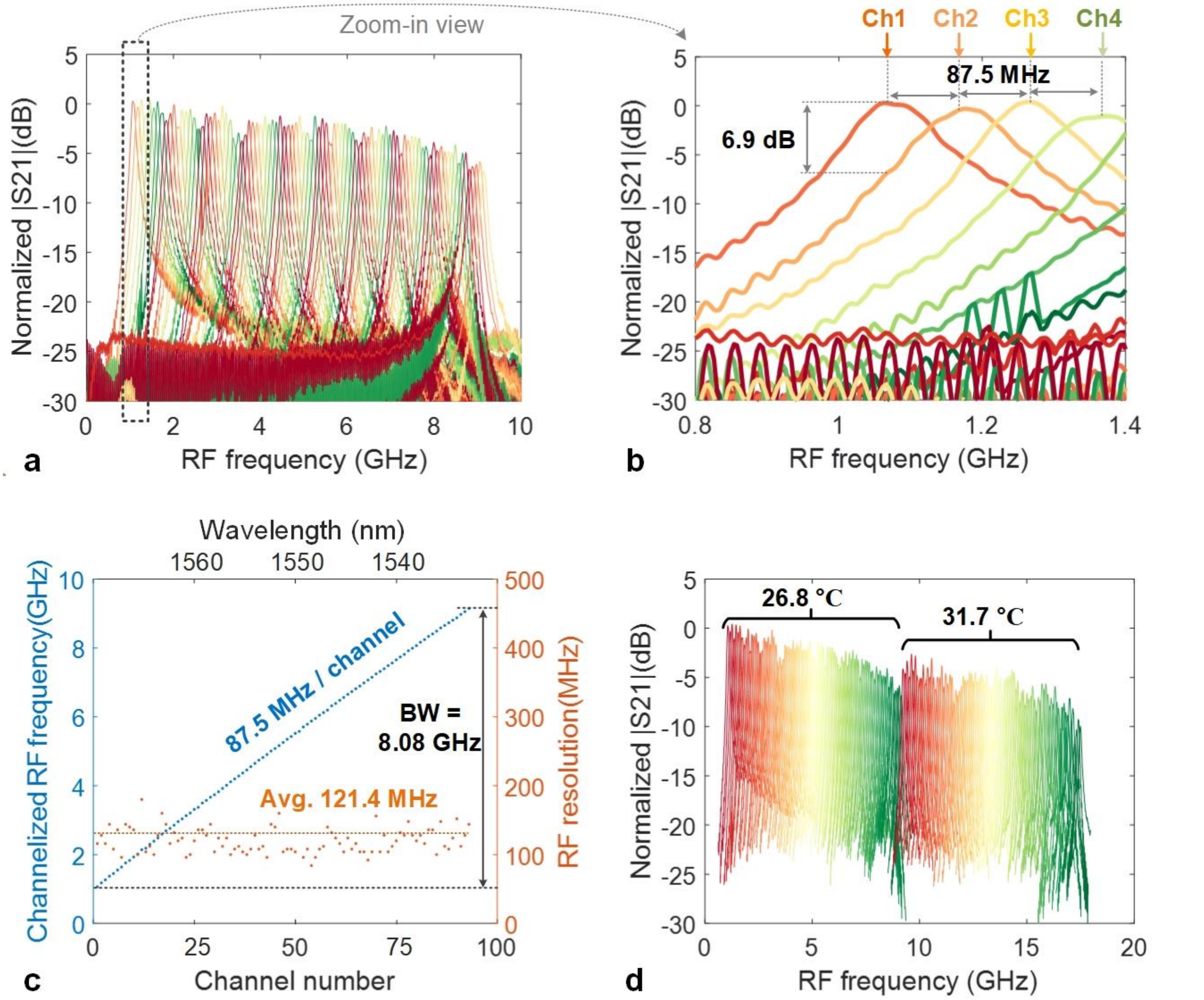

Fig. 4. Measured RF transmission spectra of (a) the 92 channels and (b) a zoom-in view of the first 4 channels. (b) Extracted channelized RF frequency and resolution. (d) Measured RF transmission spectra at different chip temperatures of the passive MRR.

We note that, here we employed phase modulation in combination with the notch filters (the through-port transmission of the passive MRR) to map the high-Q resonances of the passive MRR onto the RF domain, without the need for external local oscillator paths. The phase modulation first yielded upper and lower sidebands with opposite phases, one of which was then suppressed by the notch resonances, leaving the optical carrier and the other unsuppressed sideband beating upon photodetection, thus effectively converting the modulation format from phase modulation to intensity modulation (single-sideband). Meanwhile, 92 bandpass filters were also achieved with the bandwidth, or spectral resolution $\Delta f$, determined by the passive MRR's resonant linewidth, and the center frequencies $f_{RF}(k)$ determined by the spacing between the optical carriers and the passive resonances [$f_{MRR}(k) - f_{OFC}(k)$], as described in Eq. (2). As a result, the input RF spectrum was channelized into 92 segments, each centered at $f_{RF}(k)$ and with a bandwidth of $\Delta f$.

To quantify the key experimental parameters mentioned above, we measured the transmission spectrum of the passive MRR, as shown in Fig. 3(a), using a broadband incoherent optical source implemented through the amplified spontaneous emission of an EDFA. Both TE and TM polarized resonances were observed, as marked in the figure. As can be seen, the channelized RF frequency $f_{RF}(k)$, calculated from the spacing between the comb line and the adjacent passive resonance, progressively decreases from the blue to red. The extracted result of the channelized RF frequency (Fig. 3(b)) shows a total operation bandwidth of 8.4 GHz and 3.8 GHz for TE and TM resonances, respectively, together with a fit RF channelization step of 89.5 (TE) and 41.4 (TM) MHz per channel. Considering the channelization resolution, denoted by the linewidth of the passive resonances, is ~120 MHz (measured in the following experiments), we employed the TE passive resonances for a broader operation bandwidth and lower adjacent-channel interference.

After the passive MRR, the RF spectrum was channelized into multiple segments carried at different wavelength channels, which were de-multiplexed into parallel spatial paths and detected separately. We employed an RF Vector Network Analyzer to accurately measure the performance of our channelizer. The measured RF transmission spectra of 92 channels are shown in Fig. 4 (a, b), directly verifying the feasibility of our approach and the realization of 92 parallel RF channels. We note that the imbalanced RF power of the channels can be equalized by incorporating the RF transmission spectra into the comb shaping feedback loop, where the error signal is generated by comparing the ideal channel weights with the RF power, instead of the optical power.

The centre frequencies of each RF channel (or channelized RF frequencies), were extracted as shown in Fig. 4(c), showing a demonstrated RF channelization step of 87.5 MHz per channel and thus a total instantaneous bandwidth of 8.08 GHz—matching very well with our calculations from the MRR's transmission spectrum. The average RF resolution of the channels, calculated from the 3dB bandwidth of each channel's RF transmission spectrum, was measured to be 121.4 MHz, which was a result of the passive MRR's high Q factors. Such a high channelizing resolution greatly relaxed the requirement of ADCs bandwidth and indicates that our approach is compatible with a wide range of digital components that feature relatively small bandwidths.

We note that ideally the RF channelization step (87.5 MHz in our experiments) should equal the RF resolution (121.4 MHz), such that the instantaneous bandwidth can be maximized to 121.4MHz×92=11.17 GHz. This requires a finer adjustment of the passive MRR's FSR during the nanofabrication process. In addition, the adjacent-channel interference can be reduced by employing a high-order optical filter with a higher roll-off rate and flat passband [57].

Finally, we note that the channelizer's operation band is tunable over a wide range. By thermally tuning the passive MRR, the relative spacing between the source and filtering MRRs' resonances ($f_{MRR}(1) - f_{OFC}(1)$) can be dynamically controlled, where a millisecond time-scale thermal response time is expected. As shown in Fig. 4(d), by adjusting the chip temperature of the passive MRR, we successfully shifted the instantaneous operation band of the channelizer from 1.006-9.147 GHz to 9.227-17.49 GHz, continuously covering a total RF bandwidth of 16.48 GHz. The maximum operation frequency of our channelizer was determined by the Nyquist frequency of our microcomb source, which was 48.9 GHz/2=24.45 GHz. By employing microcombs with a larger spacing, such as 200 GHz, or perhaps by using single sideband techniques, [58,59] the maximum operation frequency could be enhanced to 50-100 GHz — although this is limited by the number of comb lines within the available wavelength range.

## IV. Conclusion

We demonstrate a broadband RF channelizer using a soliton crystal microcomb source. Due to the small (49GHz) and closely matching FSRs of the microcomb and passive MRR filters, up to 92 available wavelength channels were generated, resulting in a broad RF instantaneous bandwidth of 8.08 GHz. A high RF slicing resolution of 121.4 MHz was achieved by the high-Q passive MRR that served as the periodic optical filter for spectra slicing. Phase to intensity modulation format conversion was employed to ensure stable signal detection without the need for any external local oscillator paths. Dynamic tuning of the RF operation frequency range was achieved through thermal control applied to the passive MRR, achieving RF operation up to 17.49 GHz. This microcomb-based approach features massively parallel channels that are highly promising for broadband instantaneous signal detection and processing, representing a solid step towards fully integrated photonic receivers for modern RF systems.